# Ferroelectric Transition Induced by the Incommensurate Magnetic Ordering in $LiCuVO_4$

Yutaka Naito, Kenji Sato, Yukio Yasui, Yusuke Kobayashi,
Yoshiaki Kobayashi and Masatoshi Sato*

*Department of Physics, Division of Material Science, Nagoya University, Furo-cho, Chikusa-ku, Nagoya 464-8602*

## Abstract

A ferroelectric transition occurring simultaneously with helical spin order has been found in both the polycrystalline and single crystal samples of $LiCuVO_4$. The system has $Cu^{2+}$ spins (s=1/2) of the $CuO_2$ chains formed of edge-sharing $CuO_4$ squares. Possibly due to the frustrated nature caused by the competition between the nearest-neighbor and next-nearest-neighbor exchange interactions, Cu spins exhibit, as has been already reported, the helical magnetic order with the incommensurate modulation vector $\boldsymbol{Q}$ along the chain direction and with the helical axis $\boldsymbol{e}_3$ perpendicular to the $CuO_4$ squares. The electric polarization $\boldsymbol{P}$ can be understood by the recently predicted relation $\boldsymbol{P} \propto \boldsymbol{Q}\times\boldsymbol{e}_3$. The transition temperature has been found to be gradually suppressed by the applied magnetic field.

*corresponding author: e43247a@nucc.cc.nagoya-u.ac.jp

$LiCuVO_4$ has the orthorhombically distorted inverse spinel structure (space group *Imma*)[1] Because the edge-sharing $CuO_6$ octahedra in this structure are largely distorted by the Jahn-Teller effect, we can just see the one dimensional chains of the edge-sharing $CuO_4$ squares or the one dimensional chains of $Cu^{2+}$ spins ($s$=1/2) along the $b$ axis (Fig. 1). $VO_4$ tetrahedra with nonmagnetic $V^{5+}$ ions connect these chains. For this system, the exchange interaction between the nearest neighbor $Cu^{2+}$ ions through the Cu-O-Cu exchange paths is rather weak, as expected for the Cu-O-Cu angle close to 90° (~95°), or even weaker than the next nearest neighbor interaction,[2-4] suggesting effects of the magnetic frustration are significant in its magnetic properties. Actually, the reported magnetic structure is not trivial:[2-4] As also shown in Fig. 1, it has the helical structure with the incommensurate modulation vector $\boldsymbol{Q}\sim 0.532\boldsymbol{b}^*$ with the spins in the $CuO_4$ square planes.

This kind of magnetic structure is interesting, because it satisfies the condition for the occurrence of magnetic and ferroelectric simultaneous transitions[5,6] to the so-called multiferroic state. Moreover, the system may become, if it has the simultaneous transition, the rare example of the multiferroics with the spins only in the $x^2$-$y^2$ orbital, even though many systems with magnetic moments in the $t_{2g}$ orbitals have been reported to be multiferroic.[7-11]

In the present work, we have carried out measurements of the magnetic- and dielectric-susceptibilities on both polycrystalline and single crystal samples of $LiCuVO_4$, and found that the system really exhibits the anomalous temperature ($T$) dependences of these quantities for the applied electric field parallel to $a$, indicating that the magnetic and ferroelectric simultaneous transitions take place at the critical temperature $T_c\sim 2.4$ K. The electric polarization is expected to appear only along the $a$ axis, which is consistent with the theoretical considerations.[5,6] The magnetic field dependence of $T_c$ have also been studied.

The polycrystalline samples were prepared by the solid reaction: $LiVO_3$ and CuO were mixed with the proper molar ratio and the mixtures were pressed into pellets and sintered at 550 °C for 50 h. Single crystals of $LiCuVO_4$ were grown by a flux method[12]: The mixtures with a composition of 60 mol% of $LiVO_3$ and 40 mol% of the obtained polycrystalline $LiCuVO_4$ were held at 675 °C and then cooled to 580 °C at a cooling rate of 0.9 °C/h. The obtained crystals were checked not to have an appreciable amount of impurity phases by the powder X-ray measurements. The crystal axes were determined by observing the X-ray diffraction lines. For one of the crystals, we have checked that the incommensurate magnetic structure is realized below $T_c$ by the neutron measurements.

The magnetic susceptibilities $\chi$ were measured using a SQUID magnetometer (Quantum Design) in the temperature range from 2 K to 300 K. The dielectric susceptibility $\varepsilon$ were studied by measuring the capacitance of the thin sample plates, to which the electrodes were attached with the silver paint. The measurements were carried out by varying temperature $T$ (1.4 K$\leq T \leq$300 K) with the frequency of 1 kHz using an ac capacitance bridge (Andeen Hagerling 2500A). Because there is the stray capacitance rather large as compared with the sample capacitance, we have not determined the absolute values of the dielectric susceptibility ε. However, we could successfully detect the anomalous behavior of ε, through the possible divergent behavior of the capacitance. In the measurements, both the polycrystalline and single-crystal samples of $LiCuVO_4$ were used and for the single crystals, the electric field ***E*** was applied along various crystal axes, where only for ***E***//***a***, the ferroelectric anomaly of the dielectric susceptibility has been observed. This anomaly has also been studied under the applied magnetic fields up to 8 T.

Figure 2 shows the magnetic susceptibility χ measured with the external magnetic field $H$ =1 T for one of the polycrystalline samples of $LiCuVO_4$ with the zero-field-cooling (ZFC) condition. The peak of χ at ~30 K can be considered to be due to the growth of the short range spin correlation with decreasing $T$, and the sharp peak observed at $T_c$~2.4 K is due to the magnetic transition to the helically ordered structure.

Figure 3 shows the capacitance $C$ of the polycrystalline sample with the size of $5\times5\times0.6$ mm$^3$. We have observed a clear anomaly in the $C$-$T$ curve at 2.4 K, indicating that the ferroelectric transition takes place simultaneously with the magnetic transition. One might think that the anomaly is too small to indicate the diverging nature of the dielectric susceptibility ε of the sample. To confirm that the anomaly is due to the ferroelectric transition, we compare the present data with those of a $Ni_3V_2O_8$ single crystal with a similar size. The system is known to exhibit the ferroelectric transition simultaneously with the magnetic one.[13)] Actually, we have observed the anomaly of ε for ***E***//***b*** as shown in the inset of Fig. 3. By the comparison, we find that the anomalies observed for these two samples have similar magnitudes, indicating that the present $LiCuVO_4$ also exhibits the ferroelectric transition.

The capacitance measurements have also been made for thin plate-like single crystals of $LiCuVO_4$ with the electric fields ***E*** along all the *a*, *b* and *c* axes, and the results are shown in Fig. 4: We have not found any anomaly of the capacitance $C$ or the dielectric susceptibility ε along the *b* and *c* axes. Only for ***E//a,*** a significant anomaly has been found. These results indicate that the electric polarization ***P*** is expected to appear at $T_c$ along the *a*-axis. (The $T_c$ values of crystal samples are slightly sample dependent, as was also found in ref. 2.)

To see the magnetic field dependence of ε, we have also measured the capacitance at various magnetic fields $H$ parallel to the ***c*** axis or perpendicular to the $CuO_4$ square planes and all the results are summarized in Fig. 4. As shown in the upper inset, the temperature of the peak in the $C$-$T$ curve decreases with increasing the field $H$. It seems to be simply due to the decrease of the magnetic ordering temperature, which is usually observed for antiferromagnetic transitions. From the lower inset, we can see that the peak height $\Delta C$ exhibits the linear decrease with $H$.

So far, we have shown that $LiCuVO_4$ exhibits the transition to the multiferroic phase. Below $T_c$, the spins are, as reported previously,[2,3] in the helically ordered state and the uniform electric polarization exists along the *a* direction. To understand these results, the theories by Mostvoy[5] and Katsura *et al.*[6] can be applied. By using the order parameter expansion treatments, Mostovoy concludes that the electric polarization $\boldsymbol{P}$ can be described $\boldsymbol{P} \propto \boldsymbol{Q} \times \boldsymbol{e}_3$, where $\boldsymbol{e}_3$ is the unit vector parallel to the helical axis. Katsura *et al.*[6] calculated the polarization using a microscopic model, having the spin array with the oxygen atoms on the midpoint of two adjacent spins. They consider modulated magnetic structures such as the helical and conical ones. Then, assuming that the spins are in the $t_{2g}$ orbitals and considering the spin-orbit coupling, they derived the above relation.

In the present case, the modulation vector $\boldsymbol{Q}$ is along $\boldsymbol{b}^*$ (=0.532$\boldsymbol{b}^*$), and the helical axis $\boldsymbol{e}_3$ is perpendicular to the $CuO_4$ square planes or $\boldsymbol{e}_3 // \boldsymbol{c}$, we expect that $\boldsymbol{P}$ ($\propto \boldsymbol{Q} \times \boldsymbol{e}_3$) is along the $\boldsymbol{a}$ direction, which has been really observed in the present study. However, because their calculation is, as we stated, for the spins in the $t_{2g}$ orbitals, the situation seems to be slightly different from the present case, where the spins are only in the $x^2$-$y^2$ orbitals. If we consider only the $e_g$ orbitals, the spin-orbit coupling do not induce the mixing of the two $e_g$ orbitals and we do not expect the finite polarization within the treatments of ref. 6. (The situation of the present system has a further difference, because two oxygen atoms between the neighboring spins are not on the line connecting the spins.) Then, to derive the electric polarization, we have to consider both the $t_{2g}$ and $e_g$ orbitals. We simply expect that the polarization is much smaller than those observed for systems with spins in the $t_{2g}$ orbitals, though we have not measured the magnitude of the polarization, yet.

Apart from the model of Katsura *et al.*,[6] we can consider the possible shifts of the oxygen positions which strengthen the exchange coupling of the spins and resultingly stabilize the multiferroic phase. However, this mechanism cannot be applied to the present system, because the net polarization due to the shifts of two oxygen atoms between the two spins becomes zero, as can be found by the symmetry consideration.

In conclusion, we have found, in $LiCuVO_4$, the transition to the multiferroic state at $T_c$~2.4 K, that is, the incommensurate helical magnetic ordering and the ferroelectric polarization along the *a* direction have been found to appear simultaneously at $T_c$. It is the first example of the multiferroics with $s$=1/2 spins. It is also the first example of the spins only in the $x^2$-$y^2$ orbital. The phenomena can be understood by the theory developed by using the order-parameter expansion. It seems to be also explained by the existing microscopic model which treats the spins in the $t_{2g}$ orbitals. However, further consideration has to be made for the present spins which are in the $e_g$ orbitals.

**Acknowledgments**

The work is supported by Grants-in-Aid for Scientific Research from the Japan Society for the Promotion of Science (JSPJ) and by Grants-in-Aid on priority areas from the Ministry of Education, Culture, Sports, Science and Technology.

Figure captions

Fig. 1 Schematic structure of $LiCuVO_4$ (left). The edge-sharing chains of $CuO_4$ square planes can be seen with $VO_4$ tetrahedra connecting the chains. The Li atoms are shown by the small black circles. The magnetic structure of the $Cu^{2+}$ moments is also shown (right).

Fig. 2 Temperature dependence of the magnetic susceptibility ($\equiv M/H$) of a polycrystalline sample of $LiCuVO_4$. The inset shows the data with the enlarged scales.

Fig. 3 Temperature dependence of the capacitance measured for the polycrystalline sample of $LiCuVO_4$. The inset shows the data obtained for a single crystal sample of $Ni_3V_2O_8$ with a condition similar to that for $LiCuVO_4$. Although the contribution of the stray capacitance is rather large, we can see a significant anomaly due to the ferroelectric transition.

Fig. 4 Top figure shows the temperature dependence of the capacitance of the single crystal sample of $LiCuVO_4$ along the *a* axis at various magnetic fields *H//c*. The upper inset shows the relationship between *H* and the temperature of the peak observed in the *C*-*T* curves. In the lower inset, the peak height $\Delta C$ observed at the ferroelectric transition is shown against *H*. The results for *E* parallel to *b* and *c* are shown in the middle and bottom figures, respectively (*H*=0).

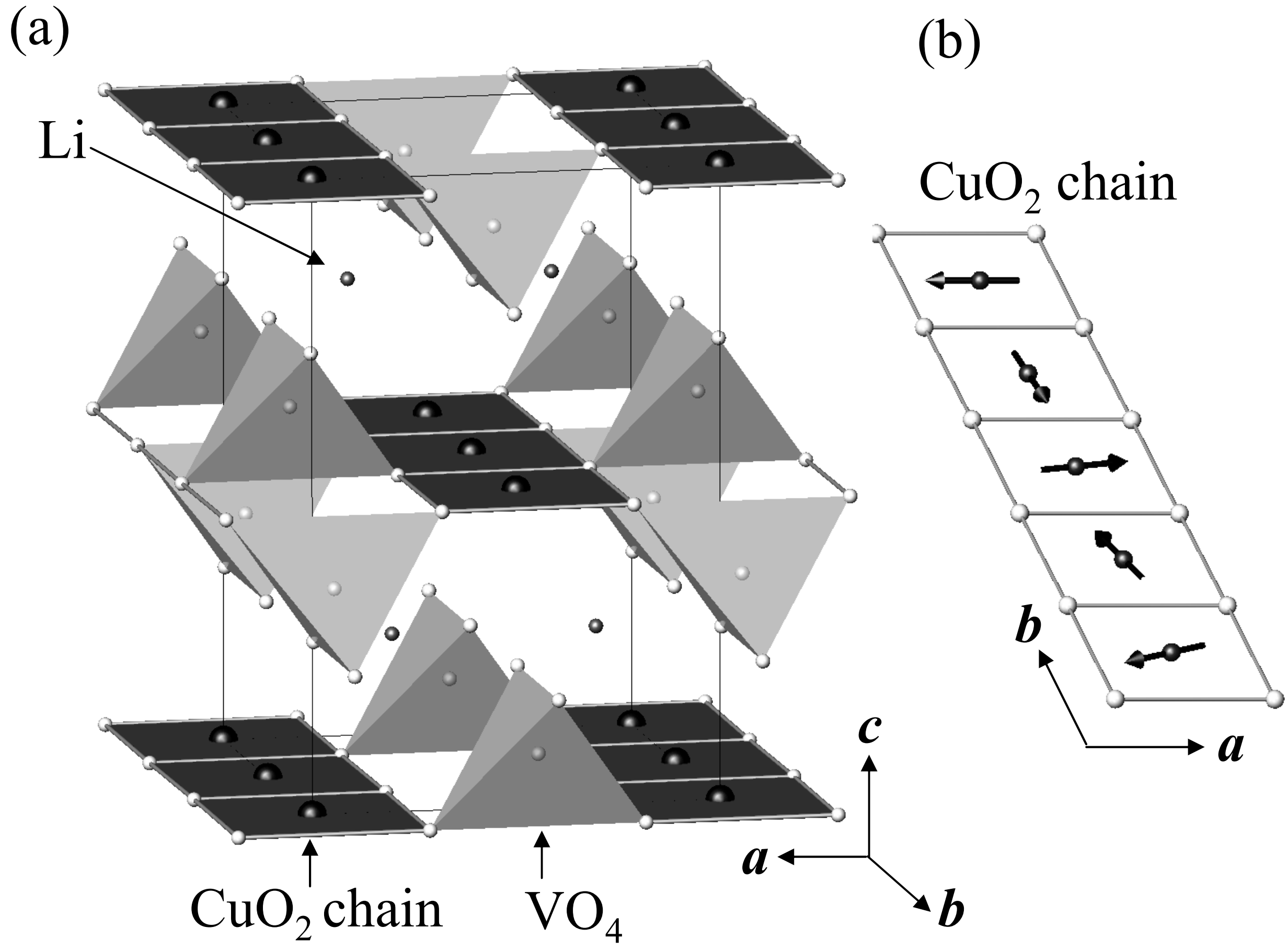

Fig. 1, Y. Naito *et al.*

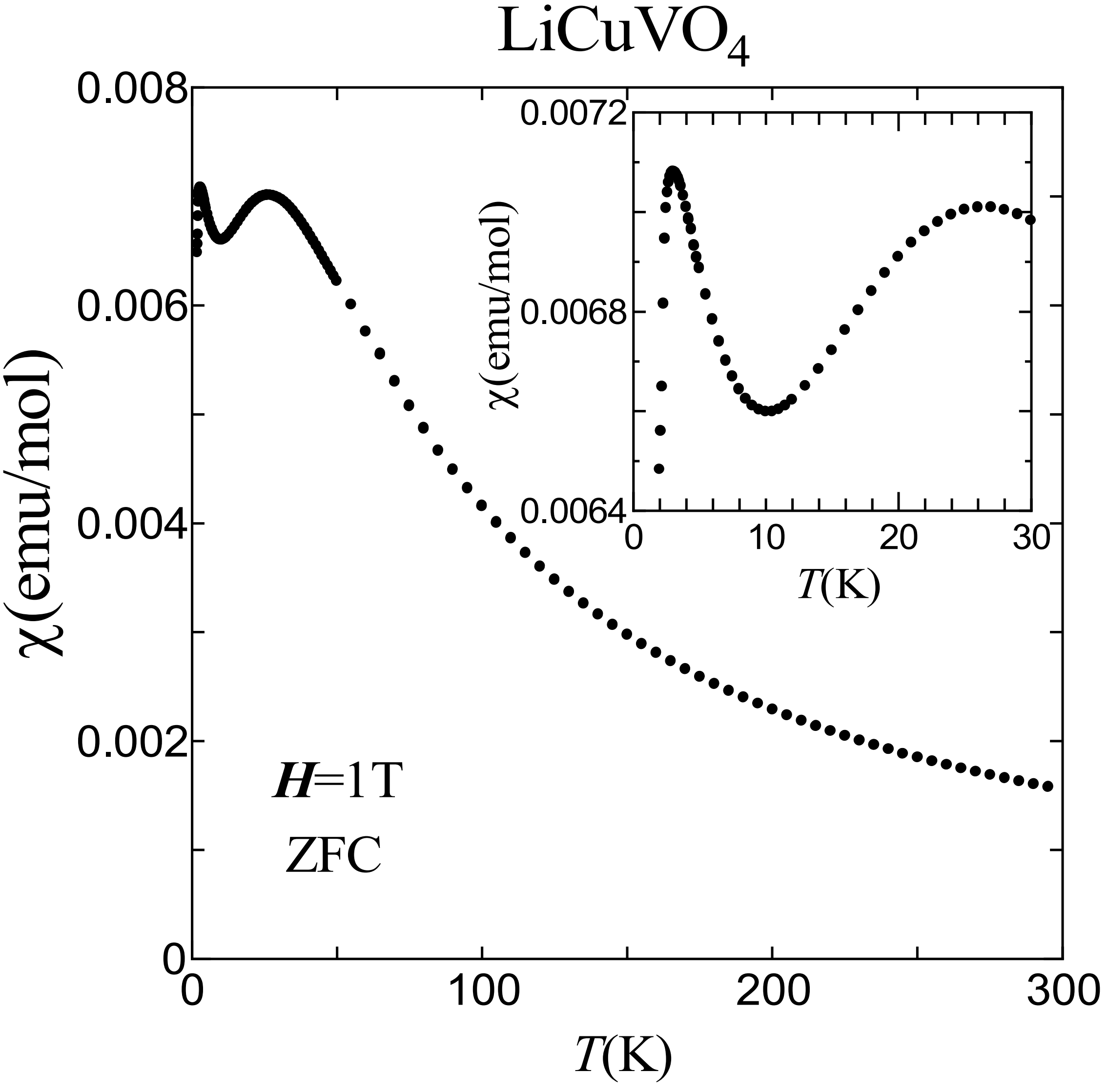

Fig. 2, Y. Naito *et al.*

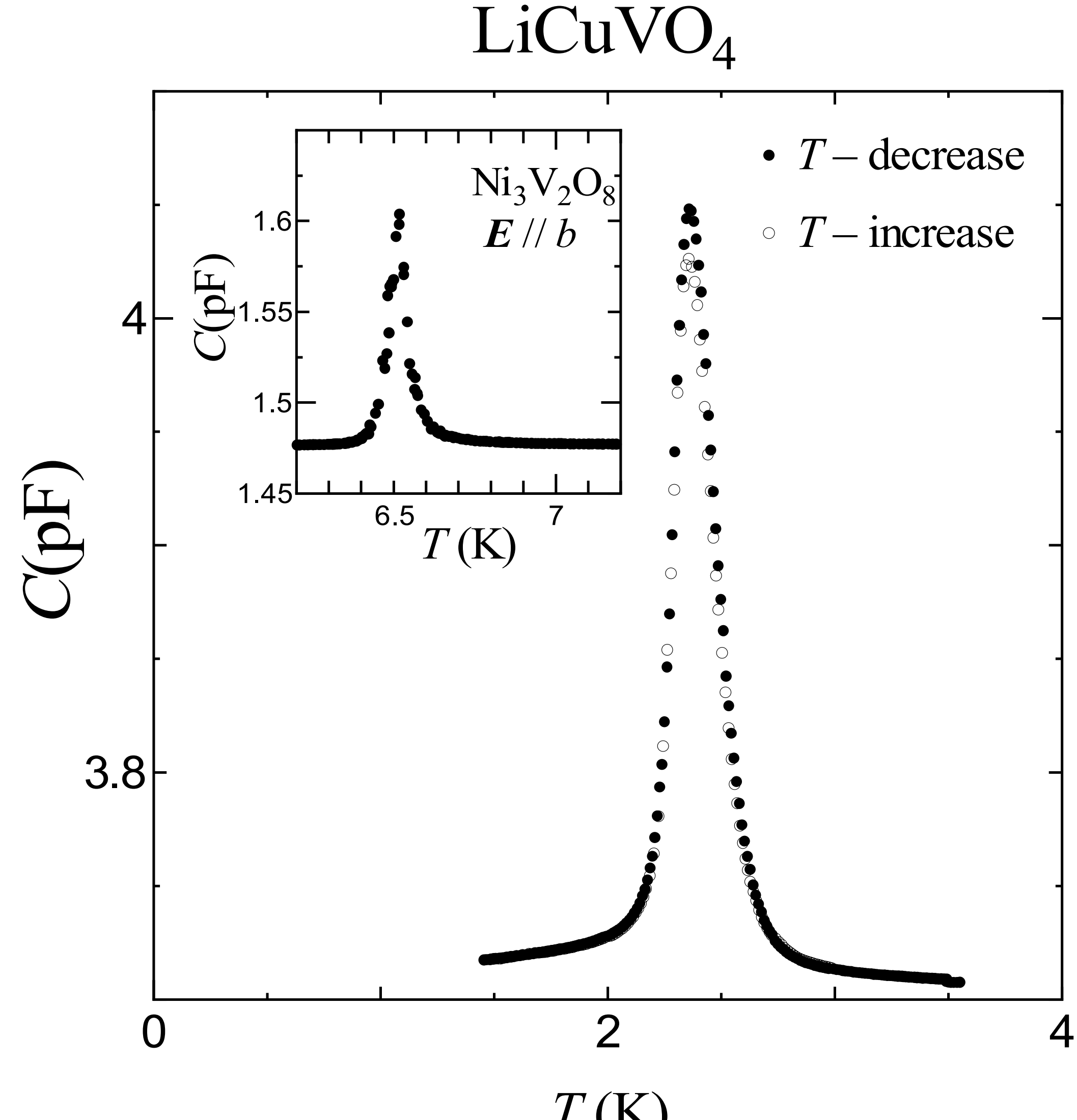

Fig. 3, Y. Naito *et al.*

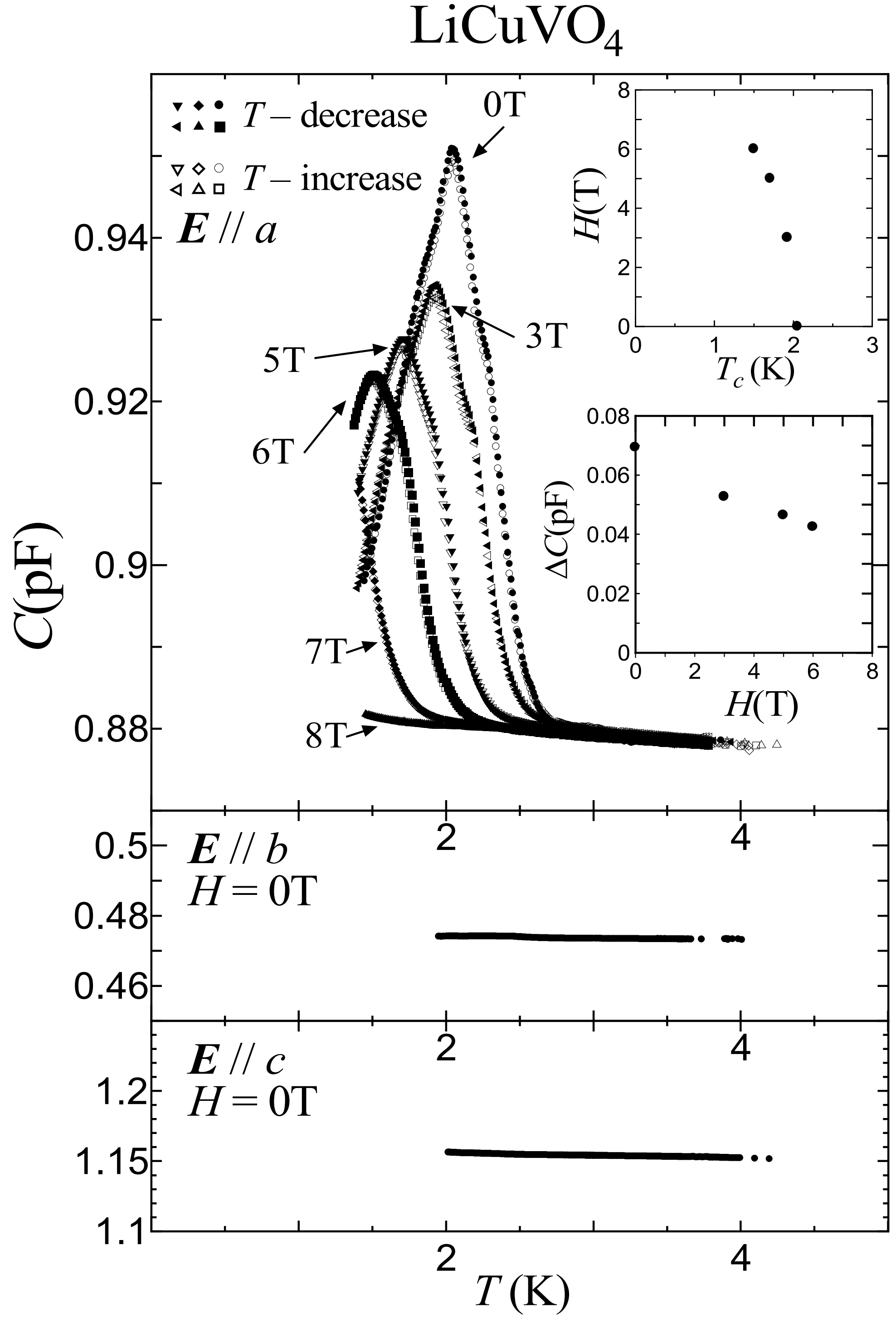

Fig. 4, Y. Naito *et al.*